\magnification=\magstep1
\tolerance 500
\rightline{TAUP 2514-98}
\rightline{5 August, 1998}
\rightline {hep-th/9808030}
\vskip 2 true cm
\centerline{\bf  Non-Orthogonality of Residues}
\centerline{\bf in the }
\centerline{\bf Wigner-Weisskopf Model for Neutral $K$ Meson Decay}
\centerline{Eli Cohen and L.P. Horwitz\footnote{*}{Also at Department
of Physics, Bar Ilan University, Ramat Gan, Israel}}
\centerline{Department of Physics and Astronomy}
\centerline{Raymond and Beverly Sackler Faculty of Exact Sciences}
\centerline{Tel Aviv University, Ramat Aviv 69978, Israel}
\vskip 2 true cm
\noindent {\it Abstract\/}: We  review  the application of the
Wigner-Weisskopf model for the
neutral $K$ meson system in the resolvent formalism. The Wigner-Weisskopf
 model is not equivalent to the Lee-Oehme-Yang-Wu
 formulation (which provides an accurate representation of the data).
  The residues in the pole approximation in the Wigner-Weisskopf model
are not orthogonal, leading to additional  interference
terms in the $K_S-K_L \,\, 2\pi$ channel.  We show that these terms
would be detectable experimentally in the decay pattern of the
 beam emitted from a regenerator if
the Wigner-Weisskopf theory were correct.
The consistency of the data with the Lee-Oehme-Yang-Wu formulation
appears to  rule out the applicability of the Wigner-Weisskopf theory
to the problem of neutral $K$ meson decay.
\vfill
\break

\par Lee, Oehme and Yang$^1$ constructed a generalization of the
Wigner-Weisskopf$^2$ decay model in pole approximation, in terms of a
two-by-two non-Hermitian effective Hamiltonian, leading to an exact semigroup
 law of evolution for the two channel $K^0$ decay.  Wu and Yang$^3$ developed,
from this model, an effective parametrization of the $K^0\rightarrow 2\pi$
decays, resulting in a phenomenology that has been very useful in describing
the experimental results. Since the early 1970's , Monte Carlo reproduction
of the data, establishing the parameter values, has been remarkably accurate
in the energy 60-120 Gev of the kaon beam$^4$.
  \par We wish to point out
 that the Wigner-Weisskopf model$^2$ is, in fact, not consistent with the
exact semigroup evolution assumed in the Lee-Oehme-Yang-Wu formulation,
corresponding to exact
 semigroup evolution, even in pole approximation,
 and that the difference between
 the Wigner-Weisskopf prediction and semigroup evolution can, in
 principle, be seen experimentally.  As we shall show, the experimental data
 rules out the applicability of the Wigner-Weisskopf
theory.
 In this letter, we treat the $K_S-K_L\,\, 2\pi$  interference channel
at the output of a regenerator. We consider
 elsewhere the structure of the two channel Wigner-Weisskopf model on the
 regeneration process itself$^5$.
   \par Regeneration systems are designed to reconstitute the $K_S$ beam
 from a $K_L$ beam in the order of $K_S:K_L \cong 10^{-3}:1$, since the $K_S$
 beam decays rapidly with high branching ratio to $2\pi$, while the $K_L$
beam decays to $2\pi$ only on the order of $10^{-3}$, the measure of
$CP$  violation.
  \par The Wigner-Weisskopf model of particle decay (we first treat
the one-channel case) assumes that there is an initial state in the
Hilbert space representation of an unstable system; in the course of time,
 Hamiltonian action evolves this state, and the component that remains
in the $\phi$ direction is called the survival amplitude:
$$   A(t) = (\phi, e^{-iHt}\phi).   \eqno(1)$$
The Hamiltonian $H$ includes a part $H_0$ for which the $K$-meson state is
stationary, and a part $H_I$ which corresponds to the part of weak interaction
leading to the decay of the $K$-meson.  We take the state $\phi$ to be a state
of the neutral $K$-meson system (e.g., some linear combination of $K^0$ and
${\overline K}^0$); this eigenstate of $H_0$ is not stable under the full
evolution $H$.
\par Defining the reduced resolvent, analytic for $z$ in the
upper half plane, in terms of the Laplace transform
$$ \eqalign{ R(z) &= -i\int_0^\infty
\,e^{izt}(\phi,e^{-iHt}\phi)dt \cr
&= (\phi, {1 \over {z-H}}\phi), \cr} \eqno(2)$$
we see that
$$ (\phi, e^{-iHt} \phi) = {1 \over 2 \pi i } \int_C dz\,
e^{-izt} \bigl(\phi, {1 \over z-H} \phi \bigr), \eqno(3) $$
where $C$ is a contour running from $\infty$ to $0$ slightly above the
real axis, and from $0$ to $\infty$ slightly below.  The lower contour
can be deformed to the negative imaginary axis, providing a
contribution
which is small except near the branch point, and the upper contour can
be deformed downward (with suitable conditions on the spectral
function) into the second Riemann sheet, where a resonance pole may
appear.  Assuming that the resonance contribution is large compared
to the ``background'' integrals, the reduced motion of the survival amplitude
is well-approximated by
$$ (\phi, e^{-iHt} \phi) \cong g e^{-iz_0t}, \eqno(4) $$
where $g$ is the $t$-independent part of the residue ($\approx$
unity), and
$$ z_0 = E_0 - i{\Gamma_0 \over 2}$$
is the pole position. It appears that this evolution could be
generalized for the two channel case by replacing $z_0$ by an
effective $2 \times 2$ non-Hermitian Hamiltonian.  As we shall see,
this structure is not generally admitted by the Wigner-Weisskopf
model.
\par For the two-channel system, consider a state $\phi_i, \,\,\,i=1,2$;
let us calculate the decay amplitude $\phi_i \rightarrow \vert
\lambda_j \rangle,\, j= 1,2$, the continuum accessible by means of the
dynamical evolution $e^{-iHt}$:
$$  \sum_j \int d\lambda \, \vert \langle \lambda_j \vert e^{-iHt}
\vert \phi_i) \vert^2 = 1 - \sum_j \vert (\phi_j , e^{-iHt}
\phi_i)\vert^2.  \eqno(6) $$
Here, the survival amplitude corresponds to the $2 \times 2$ matrix
$$ A_{ij}(t)= (\phi_i, e^{-iHt} \phi_j). $$
This amplitude can be approximated in the same way as in $(4)$ by
estimating
$$ ( \phi_i, e^{-iHt} \phi_j ) = {1 \over 2\pi i} \int_C
dz \, \bigl( \phi_i , {1 \over z-H} \phi_j \bigr)e^{-izt}.
\eqno(7)$$
\par It is convenient to write the $2 \times 2$ matrix reduced
resolvent in the form
$$ R_{ij}(z) =   \bigl( \phi_i , {1 \over z-H} \phi_j \bigr)=
 \biggl({1 \over z -
W(z)}\biggl)_{ij}, \eqno(8) $$
where $W(z)$ is a $2 \times 2$ matrix. It is almost always true (an
exception is the matrix with unity in upper right element, and all
others zero) that
$W(z)$ has the spectral decomposition
       $$ W(z) = g_1(z) w_1(z) + g_2(z) w_2(z), \eqno(9) $$
where $w_1(z)$ and $w_2(z)$ are numerical valued, and $g_1(z), g_2(z)$ are
 $2 \times 2$ matrices with the properties that
$$ \eqalign{g_1(z) g_2(z) &= 0,\cr
  g_1^2(z) = g_1(z), &\qquad g_2^2(z) = g_2(z),\cr} \eqno(10)$$
even though $W(z)$ is not Hermitian.  These matrices are constructed
from the direct product of right and left eigenvectors of $W(z)$, and
form a complete set
$$ g_1(z) + g_2(z) = 1.\eqno(11)$$
Suppose now that the short lived $K$-decay channel has a pole at
$z_S$. The matrix $W(z_S)$ (in the second Riemann sheet)
 can be represented as
     $$ W(z_S) =  z_S g_S(z_S) + z_S' g_S'(z_S), \eqno(12) $$
where $g_S g_S' =0$, and the corresponding eigenvalues are denoted by
$z_S, z_S'$. Then the reduced resolvent, in the neighborhood of $z
\cong z_S $ has the form
$$  R_{ij}(z) \cong {1 \over (z-z_S)(1-w_1'(z_S))}g_S(z_S) + {1 \over
(z-z_S')(1-w_2'(z_S))}g_S'(z_S), \eqno(13)$$
where $w_1'(z),w_2'(z)$ are the derivatives of the eigenvalues of
$W(z)$, of order the square of the weak coupling constant (these
functions correspond to the mass shifts induced by the weak interaction). The
residues are therefore $g_S(z_S)$ and $g_S'(z_S)$ to a good
approximation.
For the long-lived component, on the other hand, the pole occurs at
$z_L$, and at this point,
$$ W(z_L) = z_L g_L(z_L) + z_L' g_L'(z_L). \eqno(14)$$
The reduced resolvent in this neighborhood is then
$$  R_{ij}(z) \cong {1 \over (z-z_L)(1- w_1'(z_L))}g_L(z_L) + {1 \over
(z-z_L')(1- w_2'(z_L))}g_L'(z_L). \eqno(15)$$
Since $W(z_S)$ and $W(z_L)$ correspond to the matrix-valued function
$W(z)$ evaluated at two different points, although $g_S(z_S)$ and
$g_S'(z_S)$ are orthogonal ($g_S(z_S) g_S'(z_S) = 0 $), in general,
$g_S(z_S)$ and $g_L(z_L)$ are not.
 If there were no $CP$ violation,
the matrix would be diagonal in the $K_1,K_2$ basis, and the
two distinct idempotents would be structurally orthogonal even though they
correspond to the decomposition of the matrix at two different points.
 One can estimate the product$^{5,6}$
$$ {\rm upper}\,\, {\rm right}\,\, {\rm element}\,\,\, g_S g_L
= {\rm O}(\alpha^3),
 \eqno(16)$$
in a Lee-Friedrichs type model$^7$,
where $\alpha$ is the relative amplitude $CP$ violation (independently
of the strength of the weak interaction); the other matrix elements
are of order $\alpha^4$ or $\alpha^5$.
\par We now study the composition of the beam leaving a regenerator.
We assume that in the design of such a system, the $K_S$ component
has amplitude approximately $10^{-3}$ compared to the amplitude of the
$K_L$ beam, so that non-trivial interference may occur. We represent
the beam at the exit boundary of the regenerator as
$$\vert \psi\rangle_E = d_S\vert K_S\rangle +d_L\vert K_L\rangle,
\eqno(17)$$
where $d_S$ and $d_L$ are the amplitudes of the corresponding beam
components. We assume that the evolution of the system (in vacuum)
beyond this point is determined by the pole approximation of the two
channel Wigner-Weisskopf  reduced propagator,
$$U(\tau)=g_S e^{-i\kappa_S\tau}+g_L e^{-i\kappa_L\tau}. \eqno(18)$$
Then, the action of $U(\tau)$ on the exit state is
$$U(\tau)\vert \psi\rangle_E = \, A_S(\tau)\vert K_S
\rangle \, + \,A_L(\tau)\vert K_L\rangle,\eqno(19)$$
where
$$\eqalign{A_S(\tau) &= (d_S\langle{\tilde K}_S\vert K_S\rangle +
d_L\langle{\tilde K}_S\vert K_L\rangle) e^{-i\kappa_S \tau} \cr
A_L(\tau) &= (d_L\langle {\tilde K}_L\vert K_L\rangle + d_S\langle
{\tilde K}_L\vert K_S \rangle) e^{-i\kappa_L\tau} \cr}. \eqno(20)$$
\par It then follows that
$$ \vert \langle 2 \pi \vert U(\tau) \psi \rangle_E\vert^2= \vert
A_L\vert^2 \cdot\vert \langle 2\pi \vert K_L\rangle\vert^2 \biggl\{ 1
+ \vert{A_S \over A_L}\vert^2 \cdot \vert {1 \over \eta_{2\pi}}\vert^2
+ 2 {\rm Re}\bigl({A_S \over A_L}{1 \over \eta_{2\pi}}\bigr) \biggr\}
,\eqno(21)$$
where, as usual,
$$ \eta_{2\pi} = {\langle 2\pi \vert K_L \rangle \over \langle 2\pi
\vert K_S \rangle}. \eqno(22)$$
\par The ratio of the amplitudes $A_S$ and $A_L$ is given, from $(20)$
as
$$ \biggl({A_S \over A_L} \biggr) \cong\biggl({A_S \over A_L}
\biggr)^0\biggl[1+\bigl({d_L \over d_S}\bigr){\langle {\tilde K}_S \vert K_L
\rangle  \over \langle {\tilde K}_S \vert K_S \rangle }
\biggr],\eqno(23)$$
where we indicate the remainder for vanishing $\langle {\tilde K}_S
\vert K_L \rangle $ by a superscript zero.
 \par We see that the modification of the amplitude is not negligible.
 The ratio $d_L$ to $d_S$, as we have pointed out, should be of order
$10^3$ so that the $2\pi$ interference can be seen; on the other hand,
 $\langle {\tilde K}_S \vert K_L
\rangle $ is of order $\alpha^3$.  We estimate this quantity by noting
that the branching ratios$^8$ are $K_L \rightarrow 3\pi \sim 10^{-1},\,
K_l \rightarrow 2\pi \sim 10^{-3}-10^{-4}$ and $K_S \rightarrow 3\pi \sim
10^{-5}-10^{-7}, \, K_S \rightarrow 2\pi \sim 10^{-1}$. Hence for $K_L$, the
ratio $2\pi:3\pi$ is $10^{-2}-10^{-3}$ and for $K_S$ the ratio $3\pi :2\pi$
is $10^{-4} - 10^{-6}$. These ratios go as $\alpha^2$ (the transition matrix
elements $(3.2)$ of ref. 6, a spectral model, includes the phase space
 factors, so that the amplitudes  $CP {\rm violating}:CP {\rm conserving}
 \sim \alpha$), and we therefore take as a representative value for our
estimate $\alpha^2 \sim 10^{-3}$,
i.e., $\alpha^3$ of order $ 10^{-4}$.
 Moreover,
$\langle {\tilde K}_S \vert K_S \rangle $ is of order unity.  Hence
the additional term is of order $10^{-1}$, which, as we shall see,
could, in principle, be observed.  Substituting $(23)$ into $(21)$,
one finds that
 $$ \vert \langle 2 \pi \vert U(\tau) \psi \rangle_E\vert^2= \vert
A_L\vert^2 \cdot\vert \langle 2\pi \vert K_L\rangle\vert^2 \biggl\{ 1
+ \biggl\vert \biggl({A_S \over A_L}\biggr)^0\biggl\vert^2 \cdot
\biggl\vert {1 \over
\eta_{2\pi}'}\biggr\vert^2
+ 2 {\rm Re}\biggl(\biggl({A_S \over A_L}\biggr)^0{1 \over
\eta_{2\pi}'} \biggr) \biggr\}, \eqno(24)$$
where
$$ {1 \over \eta_{2\pi}'}= {1 \over \eta_{2 \pi}} \biggl\{1 + \biggl({d_L \over
d_S}\biggr) {\langle {\tilde K}_S \vert K_L
\rangle  \over \langle {\tilde K}_S \vert K_S \rangle }\biggr\}. \eqno(25)$$
\par This correction to $\eta_{2\pi}$ which would emerge from an
experimental study of the $2\pi$ decay would contain an approximate
$10\%$ deviation from the ratio $(22)$.  Furthermore, one may vary the
experimental conditions slightly to produce a difference in the coefficients
$d_S$ and $d_L$ by adding a thin sheet of additional regenerating
material.  Since
$$A_L \cong d_L \langle {\tilde K}_L \vert K_L \rangle e^{-i\kappa_L},
$$
and we expect $d_L$ to be relatively insensitive to such changes, the
main variations of $(24)$ are due to variations of $\zeta^{-1} =
d_S/d_L \equiv \lambda e^{i\theta}$ in $(A_S /A_L)^0$ and $\eta_{2
\pi}'^{-1}$. In particular,
$$ {\partial \over \partial \zeta} {1\over \eta_{2 \pi}'} = {1 \over
\eta_{2 \pi}} {\langle {\tilde K}_S \vert K_L \rangle \over \langle
{\tilde K}_S \vert K_S \rangle} ; \eqno(26)$$
since $\eta_{2 \pi} = {\rm O} (10^{-3})$, this derivative is ${\rm O}
(10^{-1})$. The effective $\eta_{2\pi}$ measured in this way would not
be a constant relative amplitude, but it would be quite sensitive to
variations in the regenerator structure ($\delta(1/\eta_{2 \pi}') \sim
{\rm O}(10^{-1})\delta\zeta $).
 The sensitivity to such changes can be seen by casting $(24)$
into a form in which the phases and real amplitudes are explicit:
 $$ \eqalign{\vert \langle 2 \pi \vert U(\tau) \psi \rangle_E\vert^2 =
  \vert A_L\vert^2 \cdot\vert \langle 2\pi
 \vert K_L\rangle\vert^2 \biggl\{ 1
&+   ({b \over 2})^2\biggl( 1 + {a \over b} \bigl[ {a \over b} + 2
\cos (\theta_{{\tilde S} L} - \xi_S - \theta) \bigr] \biggr) + \cr
&+ \sqrt{{\overline a}^2 + {\overline b}^2 } \sin (\gamma +\tau m_{LS})
\biggr\}, \cr } \eqno(27)$$
where we have used the definitions:
$$\eqalign{ \langle {\tilde K}_S \vert K_L \rangle &= K_{{\tilde S} L}e
^{i\theta_{{\tilde S}L}} \cr
a = {2 K_{{\tilde S}L} \over \vert \eta_{2 \pi} \vert m_L} e^{{1\over
2} \tau \Gamma_{LS}} &\qquad b = {2 \lambda m_S \over \vert \eta_{2
\pi} \vert m_L } e^{ { 1\over 2} \tau \Gamma_{LS}} \cr
\Gamma_{LS} = \Gamma_L - \Gamma_S &\qquad m_{LS} = m_L - m_S \cr
\kappa_{L,S} &= m_{L,S} - i {\Gamma_{L,S}\over 2} \cr
{d_S \over d_L} &= \lambda e^{i\theta} \cr \eta_{2 \pi}= \vert \eta_{2
\pi}\vert e^{i\phi} &\qquad \eta_{2 \pi}'= \vert \eta_{2
\pi}'\vert e^{i\phi'} \cr
\langle {\tilde K}_S \vert K_S \rangle = \mu_S e^{i\xi_S} &\qquad
 \langle {\tilde K}_L \vert K_L \rangle = \mu_L e^{i\xi_L} \cr
{\overline a} = a \cos(\theta_{{\tilde S} L} - \xi_L - \phi) &+ b \cos
(\xi_S - \xi_L + \theta - \phi) \cr
{\overline b} = -[ a \sin (\theta_{{\tilde S}L} - \xi_L - \phi) &+ b
\sin ( \xi_S - \xi_L + \theta - \phi) ] \cr } \eqno(28)$$
and
$$ \gamma = \arctan \bigl({ {\overline a} \over {\overline b}}\bigr).
 \eqno(29)$$
\par A value of $\eta_{2\pi}'$ obtained in this way which is stable
under variation of the regenerator structure would rule out the
applicability  of the Wigner-Weisskopf model to two channel particle
decays as exemplified in the neutral K meson system. The observed
stability of $\eta_{2\pi}$ under a wide variation of regenerator
configurations$^4$ is consistent with the Lee-Oehme-Yang-Wu parametrization,
and we therefore conclude that the Wigner-Weisskopf model is not applicable to
the description of the decay of the neutral $K$-meson system.
\bigskip
\noindent
{\it Acknowledgements}
\par We wish to thank I. Dunietz, C. Newton, B. Winstein, and T.T. Wu for
helpful discussions.
\bigskip
\frenchspacing
\noindent
{\it References}
\smallskip
\item{1.} T.D. Lee, R. Oehme and  C.N. Yang, Phys. Rev. {\bf 106}, 340
(1957).
\item{2.} V.F. Weisskopf and E.P. Wigner, Zeits. f. Phys. {\bf 63}, 54 (1930);
{\bf 65}, 18 (1930).
\item{3.} T.T. Wu and C.N. Yang, Phys. Rev. Lett. {\bf 13}, 380
(1964).
\item{4.} B. Winstein, {\it et al}, {\it Results from the Neutral Kaon
Program at Fermilab's Meson Center Beamline, 1985-1997,\/}
FERMILAB-Pub-97/087-E, published on behalf of the E731, E773 and E799
Collaborations, Fermi National Accelerator Laboratory, P.O. Box 500,
Batavia, Illinois 60510.
\item{5.} E. Cohen and L.P. Horwitz, in preparation.
\item{6.} L.P. Horwitz and L. Mizrachi, Nuovo Cim. {\bf 21A}, 625
(1974).
\item{7.} K.O. Friedrichs, Comm. Pure Appl. Math. {\bf 1}, 361 (1948);
T.D. Lee, Phys. Rev. {\bf 95}, 1329 (1956).
\item{8.} {\it Reviews of Particle Physics}, Phys. Rev. {\bf D54}, 1 (1996).

\vfill
\end
\bye